\documentclass[10pt,letterpaper]{article}
\usepackage[top=0.85in,left=2.75in,footskip=0.75in]{geometry}

\usepackage{amsmath,amssymb}

\usepackage{changepage}

\usepackage[utf8x]{inputenc}

\usepackage{textcomp,marvosym}

\usepackage{cite}

\usepackage[right]{lineno}
\usepackage{float}
\usepackage{microtype}
\DisableLigatures[f]{encoding = *, family = * }

\usepackage[table]{xcolor}

\usepackage{array}

\newcolumntype{+}{!{\vrule width 2pt}}

\newlength\savedwidth

\raggedright
\usepackage[aboveskip=1pt,labelfont=bf,labelsep=period,justification=raggedright,singlelinecheck=off]{caption}

\makeatletter
\renewcommand{\@biblabel}[1]{\quad#1.}
\makeatother

\usepackage{lastpage,fancyhdr,graphicx}
\usepackage{epstopdf}
\fancyheadoffset[L]{2.25in}
\fancyfootoffset[L]{2.25in}
\begin{document}
\vspace*{0.2in}
\begin{flushleft}
{\Large
\textbf\newline{Hiring in the substance use disorder treatment related sector during the first five years of Medicaid expansion} 
}
\newline

Olga Scrivner\textsuperscript{1\textsection*},
Thuy Nguyen\textsuperscript{2\textsection},
Kosali Simon\textsuperscript{2,3\textsection},
Esm\'e Middaugh\textsuperscript{1\ddag},
Bledi Taska\textsuperscript{4\ddag},
Katy B\"orner\textsuperscript{1\ddag},

\bigskip
\textbf{1} School of Informatics, Computing, and Engineering, Indiana University, Bloomington, IN, USA
\\
\textbf{2} O’Neill School of Public and Environmental Affairs, Indiana University, Bloomington, IN, USA
\\
\textbf{3} National Bureau of Economic Research, Cambridge, Massachusetts, USA
\\
\textbf{4} Burning Glass Technologies, Boston, Massachusetts, USA
\\
\bigskip

\textsection These authors contributed equally to this work.

\ddag These authors also contributed equally to this work.

* Corresponding author \\
obscrivn@indiana.edu (OS)

\end{flushleft}
\section*{Abstract}
Background: Effective treatment strategies exist for substance use disorder (SUD), however severe hurdles remain in ensuring adequacy of the SUD treatment (SUDT) workforce as well as improving SUDT affordability, access and stigma. Although evidence shows recent increases in SUD medication access from expanding Medicaid availability under the Affordable Care Act, it is yet unknown whether these policies also led to a growth in the changes in the nature of hiring in SUDT related workforce, partly due to poor data availability. Our study uses novel data to shed light on recent trends in a fast-evolving and policy-relevant labor market, and contributes to understanding the current SUDT related workforce  and the effect of Medicaid expansion on hiring attempts in this sector.

Methods and Data: We examine attempts over 2010-2018 at hiring in the SUDT and related behavioral health sector as background for estimating the causal effect of the 2014-and-beyond state Medicaid expansion on these outcomes through ``difference-in-difference'' econometric models. We use Burning Glass Technologies (BGT) data covering virtually all U.S. job postings by employers. 

Findings: Nationally, we find little growth in the sector’s hiring attempts in 2010-2018 relative to the rest of the economy or to health care as a whole. However, this masks diverging trends in subsectors, which saw reduction in hospital based hiring attempts, increases towards outpatient facilities, and changes in occupational hiring demand shifting from medical personnel towards counselors and social workers. Medicaid expansion did not lead to any statistically significant or meaningful change in overall hiring attempts in the SUDT related sector during this time period, although there is evidence of increases among some occupations.
	
Conclusions: Although nationally, hiring attempts in the SUDT related sector as measured by the number of job advertisements have not grown substantially, there was a shift in the hiring landscape. Many national factors including reimbursement policy may play a role in incentivizing demand for the SUDT related workforce, but our research does not show that recent state expansions in Medicaid was one such factor. Future research is needed to understand how aggregate labor demand signals translate into actual increases in SUDT workforce and availability.

\section*{Introduction}

Worldwide, the burden of opioid dependence increased by 74\% between 1990 and 2010 and has become the largest contributor to global disease burdens attributable to drug misuse in 2010 (9.2 million disability-adjusted life years). This is due to the substantial role opioids play in premature mortality, high disability, and the relatively large population with substance use disorders (SUDs)~\cite{Degenhardt2013}. The monetized burden of prescription opioid misuse alone in the United States is estimated to be \$78.5 billion a year~\cite{Florence2016a}, while at least 2.3 million Americans suffer from SUDs due to opioids~\cite{Haffajee2018}. The most effective SUD  treatment (SUDT) is a combination of long-acting medications (usually methadone or buprenorphine) administered as part of a cognitive behavioral approach (such as counseling, family therapy, and peer support programs)~\cite{Schuckit2016}.  The National Survey of Substance Abuse Treatment Services (NSSATS) reports that in 2017 there were 13,857 treatment facilities in the US with over 1,356,015 clients enrolled, representing only a 19\% increase in total clients served since 2007~\cite{SubstanceAbuseMentalHealthServicesAdministration2017}. 

The SUDT workforce is deemed inadequate by almost any measure~\cite{Saloner2015,Hoge2013}. Workforce shortages and barriers have played a prominent role in limiting treatment access among those suffering from SUDs~\cite{Haffajee2018,Bouchery2018}. The needs of those suffering from SUDs is also broader than just addictions treatment, as mental health is a very frequent co-occurring or resulting need~\cite{Flynn2008}. 

Thus, our empirical focus in this paper due to clinical evidence and due to data limitations discussed later, is the "mental health and substance abuse treatment" workforce as classified by the North American Industrial Classification System, which we refer to as the SUDT and related workforce or sector throughout this paper.

Substantial resources are being allocated to address the current US opioid crisis, in part by increasing the pipeline of the SUDT and related workforce and increasing access to medications~\cite{DepartmentofHealthandHumanServices}, yet less than 20\% of patients with SUDs received any treatment in 2009-2013~\cite{Saloner2015}.  Methadone is provided to American patients only through licensed, accredited and closely monitored clinics, opioid treatment programs (OTPs) which are stably about 10\% of the specialty treatment systems~\cite{Schuckit2016,Mojtabai2019}. Despite increasing demand and perpetual waitlists for treatment, the supply of OTPs has remained low and constant over time, with around 1,500 approved programs in 2017 compared to 1,166 OTP programs reported in 2010~\cite{Haffajee2018}. Alternatively, SUD patients can receive buprenorphine maintenance therapy from office-based providers (physicians, nurses practioners and physicians' assistants) approved to prescribe buprenorphine~\cite{Schuckit2016}. Lack of buprenorphine-waivered providers is prevalent; in 2016 no buprenorphine waivered providers were found in 47\% of all US counties, and in 72\% of rural counties~\cite{Christie2017}. Persistent workforce barriers, leading to treatment underutilization, include insufficient education and training, burdensome regulatory procedures, lack of ability to refer patients for mental health and substance abuse counseling, burdensome reimbursement barriers, and provider stigma~\cite{Haffajee2018}.

In the 36 states and DC, where Medicaid has expanded through the Affordable Care Act (ACA), Medicaid insurance inclusion has broadened to practically all adults with income levels beneath the benchmark of 138 percent of the federal poverty level~\cite{KaiserFamilyFoundation2019}. Prior evidence suggests that Medicaid expansion has led to substantial increases in Medicaid reimbursement for OUD~\cite{Wen2017,Saloner2018,Maclean2019,Cicero2017a,Sharp2018}.  Particularly, evaluating the Medicaid State Drug Utilization Database (SDUD) from 2011 to 2014, Wen et al. established that a 70 percent increase in buprenorphine prescribing and 50 percent rise in associated spending had arisen as a result of Medicaid expansions; it is not yet known how these translate to increases in total use of buprenorphine as some may use have been uncompensated prior to Medicaid expansion~\cite{Wen2017}. Sharp et al. show that while Medicaid expansion resulted in reduced methadone utilization, both buprenorphine and naltrexone prescriptions increased, as exhibited by the 2011-2016 SDUD data~\cite{Sharp2018}. State Medicaid programs also facilitates access to inpatient and outpatient treatment services such as institutions for mental disease, inpatient and outpatient detoxification, psychotherapy, peer support, supported employment, and partial hospitalization~\cite{Orgera2019}. 

As the medical and service use of OUD treatments has increases following Medicaid expansion, and sources of financing now exists for comprehensive treatment of non Rx forms as well, these increases may lead to major, yet uncertain implications for mental health and addiction workforce demand~\cite{Hoge2013}. This paper examines the impacts of Medicaid expansion on job openings in the SUDT and related health care sector and investigates the nature of hiring in terms of occupations sought using data on the near-universe of 2010-2018 online US job vacancies collected by Burning Glass Technologies (henceforth BGT). While BGT has proved useful in the labor economics literature to study the effects of major policies, such as state minimum wage laws on labor demand~\cite{Clemens2018}, it has thus far not been used to study the SUDT workforce. BGT represents a valuable resource for this topic since typical labor data sets such as Bureau of Labor Statistics (BLS) products are not available with less than 2-3 year lags, yet the addictions crisis is fast moving, and the BGT data we use extends to the end of 2018, allowing us to examine recent trends in the sector.

Our approach takes advantage of standard difference-in-difference (DD) designs used in Medicaid expansion literature by comparing in this case, job postings between Medicaid expansion and non-expansion states, before and after expansion, to test the hypothesis that insurance availability increases hiring attempts in the SUDT related sector. Specifically, we extract SUDT-related job openings and aggregate data to the state level. We then compare the number of online job postings in Medicaid expansion and non-expansion states from 2010 through 2018, testing for changes in the relationship post expansion dates.  The findings from this exercise provide evidence on a large area of recent insurance policy on hiring attempts in the SUDT related workforce and has major implications regarding ability to affect SUD treatment access.

\section*{Materials and methods}
\subsection*{Datasets}

\subsubsection*{Medicaid}
Our analyses center on comparison between states that expanded Medicaid by the end of our study period (33 by 2018) versus non-expansion states (18).   Medicaid expansion status information comes from the Kaiser Family Foundation~\cite{KaiserFamilyFoundation2019}.

The 33 expansion states are: AK, AZ, AR, CA, CO, CT, DE, DC, HI, IL, IN, IA, KY, LA, MD, MA, MI, MN, MT, NH, NJ, NM, NY, ND, OH, OR, PA, RI, VT, WA, WV, and WI. All listed states had expanded Medicaid through the ACA in the first quarter of 2014 with the following exceptions: Michigan (expanded April, 2014), New Hampshire (August, 2014), Pennsylvania (January, 2015), Indiana (February, 2015), Alaska (September, 2015), Montana (January, 2016), Louisiana (July, 2016), and Wisconsin (had not formally expanded by 2018). The late expansion states were excluded in our simple mean comparison. In the regression analyses, these states were included with the actual year of expansion. In partial implementation years, the treatment status is coded as a fraction of actual months over 12 months. For instance, the treatment status for Michigan equals 3/12 in 2014, and equals 1 in 2015 and the following years. Though not actually adopting ACA expansion, we consider Wisconsin an expansion state due to its Medicaid coverage of adults up to the federal poverty threshold income level. 

The 18 non-expansion states are: AL, FL, GA, ID, KA, ME, MS, MO, NC, NE, OK, SC, SD, TN, TX, UT, VA, and WY. Medicaid expansion has been authorized for implementation in 2019 or later in five of these non-expansion states (VA, ME, ID, NE, and UT); we treat them as non-expansion states as our data period ends in 2018.

\subsubsection*{Hiring activity}
Our primary outcome, attempted hiring activity by employers, comes from a database of online job postings curated by BGT, a labor market analytics company that scrapes, cleans, and parses online job advertisements from approximately 40,000 job boards and websites~\cite{Deming2018}. The BGT data include industry and occupation codes, geographical location, and time of job postings, among other job identifiers. In this study, we focus on the time frame between 2010 and 2018, resulting in 174 million U.S. online job vacancies across all sectors of the economy. Our main outcome of interest is the hiring activity in all SUDT related establishments. According to the Substance Abuse and Mental Health Service Administration (SAMHSA), SUDT establishments are defined by the type of care offered and include outpatient, residential (non-hospital), and hospital inpatient services~\cite{SubstanceAbuseandMentalHealthServicesAdministration2018a}. Outpatient centers may provide ambulatory detoxification, methadone/buprenorphine maintenance, or naltrexone treatment; residential facilities may provide short- or long-term care as well as detoxification; and hospitals may offer medically-controlled and monitored inpatient detoxification and treatment. Emergency rooms, private doctors’ offices, self-help groups, prison and jails are not considered treatment facilities by this definition~\cite{SubstanceAbuseandMentalHealthServicesAdministration2018a}. We used 6-digit North American Industry Classification System (NAICS) codes to identify the aforementioned SUDT related establishments as follows: (i) Psychiatric and Substance Abuse Hospitals (622210), (ii) Outpatient Mental Health and Substance Abuse Centers (621420), and (iii) Residential Mental Health and Substance Abuse Facilities (623220). Thus, it is not possible in standard industry classifications to separate mental health and substance disorder treatment facilities. The BGT database records there being 48,587 job vacancy postings belonging to these three SUDT related establishments (henceforth SUDT hospitals, outpatient SUDT centers, and residential SUDT facilities, respectively, for brevity) over the 2010-2018 period. Our analysis also explores trends in hiring by occupation, given that there are quality implications on SUD patient care that depend on the composition of the workforce. BGT classifies each job vacancy by 6-digit NAICS code and by Standard Occupational Classification System (SOC) code, which enables us to document the level of hiring activities per specific occupation sought in the ads. 
\subsubsection*{Covariates}
We control for important state characteristics that may be associated with SUD-related labor market demand and be inadvertently causally attributed to Medicaid policy: unemployment rates, state populations, median household income, opioid prescribing rates, and drug poisoning mortality rates. Data on unemployment rates are from the Bureau of Labor Statistics (BLS). State population estimates and median household incomes come from the Census Bureau. The opioid prescribing rates, measured as retail opioid prescriptions dispensed per 100 persons per year, come from the Centers for Disease Control and Prevention (CDC)~\cite{CDC2019}. Drug poisoning mortality rates come from the National Center for Health Statistics and refer to the estimated age-adjusted mortality rates; they reflect the average number of drug poisoning deaths per 100,000 persons~\cite{RossenLMBastianBWarnerMKhanD2019}.
\subsection*{Descriptive information}

\begin{figure}[H]
\includegraphics[width=5.3in]{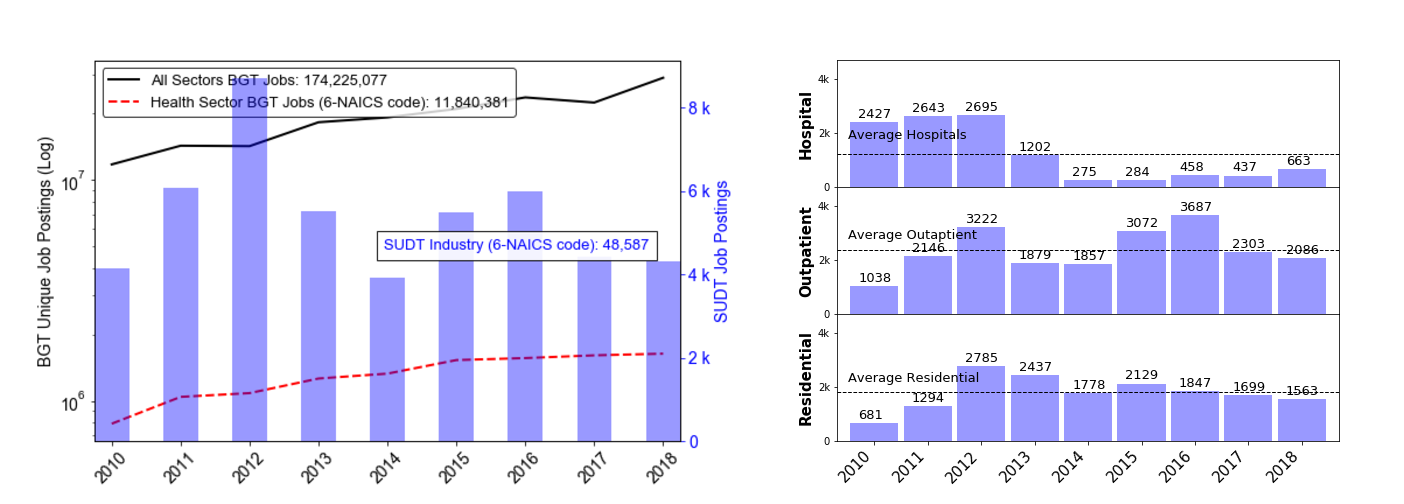}
\caption{{\bf BGT Online Job postings.} (A) BGT Job postings for All industries (black), Healthcare Industry (red) and industries (black), Healthcare Industry (red) and SUD industries (light blue). The aggregated amount for all job postings is calculated for the period from 2010 through 2018. The healthcare sector is identified by the NAICS code ‘62’. The SUDT facilities are identified by three NAICS codes ‘622210’, ‘621420’,’623220’. The left y-axis corresponds to the Logarithmic trend lines for the total of all BGT job vacancies (black solid line) and the total of BGT healthcare sector (red dashed line).  The y-right axis represents the SUDT sector values, shown as bar graphs. (B)  Break down of Rate of Change for Three SUDT sectors. Three  SUDT sectors are  represented by their number of annual job postings. Average line is calculated for each  SUDT  sector.  Data Source: Burning Glass Technologies. 2019.}
\label{fig1}
\end{figure}

Health Care and Social Assistance, classified as NAICS industry sector 62, represents 14\% of the labor force~\cite{BureauofLaborStatistics2019}; this comes to 21 million of the nearly 156 million in the labor force as of 2018~\cite{BureauofLaborStatistics2019,BureauofLaborStatistics2019a}.  Given our main focus on the hiring aspect of labor force, we first assess existing estimates of employer demand from standard national-level BLS data -- Job Openings and Labor Turnover Survey (JOLTS)\footnote{The publicly accessible JOLTS data is not available at sub national nor for 6-digit NAICS industry classification.}. The healthcare sector represents nearly 14 million job postings 
in 2018 and about 1 million jobs a month on average~\cite{Carnevale2014,Stubbs2017}. It should be noted that JOLTS measures active job postings, that is, the same posting will also be counted in the consecutive months if the position is not filled. In contrast, BGT measures true new postings: if the same advertisement occurs in the consecutive months, it will not be counted twice.  The healthcare sector (62) represents 4.7 million job ads in BGT in 2018.  
This sector constitutes about 6.8\% (11,840,381 job ads) of all the 2010-2018 BGT data (174,225,077 job ads). Job ads of three SUDT industries comprise approximately 0.5\% of the BGT health industry, including 21,290 job ads of outpatient SUDT centers, 16,213 job ads of residential SUDT facilities, and 11,084 job ads of SUDT hospitals. Of the SUDT sector, outpatient centers, residential facilities and hospitals thus make up 45\%, 33\% and 22\% of the job ads, respectively. This is consistent with the NSSAT 2017 survey, in which outpatient programs outnumber other treatment facilities~\cite{SubstanceAbuseandMentalHealthServicesAdministration2018a}. While the general trend for all as well as health industry specific ads is uniformly upward during 2010-2018, the SUDT industry exhibits a sharp decrease after 2012 Fig~\ref{fig1}A. Prior studies noted a decline only for publicly-owned SUDT facilities (17\%), as compared to the increase (19\%) for private for-profit facilities~\cite{Cummings2016}. Within the SUDT sector, Fig1A shows the SUDT hospitals hiring patterns diverge from outpatient SUDT centers and residential SUDT facilities, with hospitals showing a clear decline in job vacancies after 2012. Outpatient SUDT centers and residential SUDT facilities are similar in their trends with the exception of 2015 and 2016, in which outpatient SUDT centers show a higher increase in job advertisements (see Fig~\ref{fig1}B). To compare these trends with the actual number of SUDT establishments, we extracted data from the County Business Patterns (CBP) for each of the SUDT industry codes. Between 2010 and 2016, outpatient SUDT centers increased on net by 1,789 to reach a total of 10,967 centers across U.S. by 2016, residential SUDT facilities increased by 1,006 to total 7,943 establishments, whereas SUDT hospitals only added 11 new establishments nationally, to reach a total of 663 in 2016.

\begin{figure}[H]
\includegraphics[width=5in]{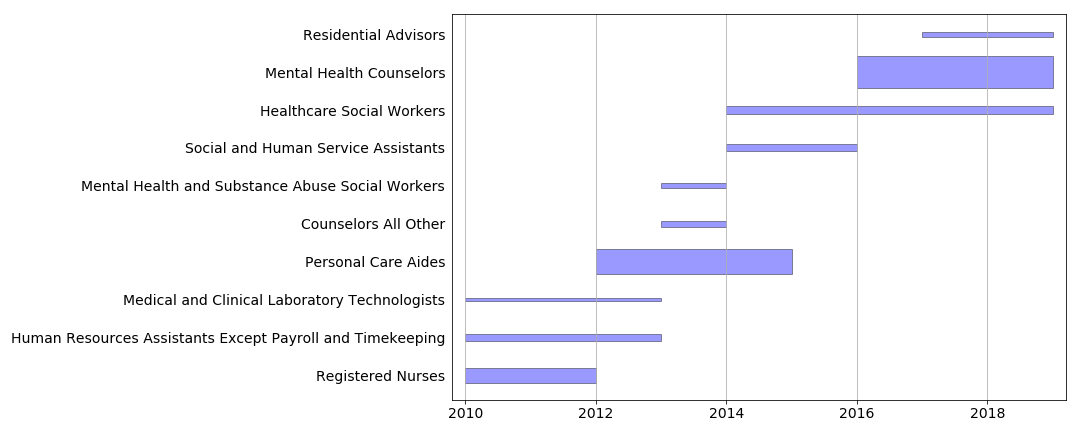}
\caption{{\bf Bursting top-10 Occupations in SUDT during 2010-2018.}
Each spike in demand is shown as a horizontal bar with a start and an end date. The length of the bar corresponds to the duration of the hiring burst, the width of the bar shows the burst strength, measured as weight. The Mental Health Counselor occupation has the strongest burst in the years 2016-2018, while the longest burst is shown for healthcare Social Workers staring from 2014 onwards.}
\label{fig2}
\end{figure}

The BGT data is also unique in allowing to track occupations specific to our 6-digit NAICS industry codes. Our analysis of specific occupations in the SUDT-related postings yielded 457 unique SOC occupations.  Occupation is listed in the vast majority (97\% or 47,592) of all SUD job postings; only 3\% (1,522) are unclassified occupation job postings. Among the 48,345 occupation-specified job ads, we have identified the following 5 most frequent occupations in this order: registered nurses, medical and health service managers, mental health counselors, personal care aides, and nurse practitioners. To detect any sudden increases in hiring activities, we perform a Kleinberg burst detection algorithm, a technique often used to identify unusual activities in events or novelty in terms~\cite{Kleinberg2003,Sci2Team2009}. Out of 457 occupation titles across 2010--2018, a total of 165 occupations displayed sudden spikes in demand and 11 occupations show double burst, for example among Training and Development Specialists there was an increased demand in 2010 and in 2018.  The top-5 bursting occupations (spikes in demands), however, occur for mental health counselors, personal care aids, registered nurses, healthcare social workers, and social and human service assistants (see Fig~\ref{fig2}). Furthermore, the burst events show a shift from registered nurses, personal care aides to mental health and healthcare social workers.

\subsection*{Causal Analysis: Methods for estimating impact of Medicaid expansion}
Our DD method essentially compares the average frequency of online US job vacancies for three SUDT-related industries and 457 SUDT unique occupations, in Medicaid expansion and non-expansion states, after policy change vs before. In order to comprehensively examine the effects of Medicaid expansion on hiring attempts by occupation, we grouped various SOC occupations into: (i) behavioral health professions including psychiatrists and psychologists, social workers, counselors, and therapists~\cite{SubstanceAbuseandMentalHealthServicesAdministration2018a}; (ii) entry-level practitioners such as personal care aides, residential advisors, social and human service assistants, nursing assistants, and home health aides; (iii) mid-level practitioners including physician assistants, nurse practitioners, registered nurses, and clinical laboratory technicians/technologists; (iv) Advance-level, primary practitioners including physicians and surgeons. These 4 groups represented 54.8 percent of all SUDT job vacancies during 2010-2018. 
To identify any causal effects of Medicaid expansion on SUD-related job vacancies, we draw on variation across states in adoptions of Medicaid expansion in a DD empirical design. Specifically, in the Ordinary Least Squares (OLS) models, we control for: (i) state fixed effects, (ii) year fixed effects, (iii) time-variant demographic factors including unemployment rates (\%) and median household income (logged), (iv) time-variant SUD-related characteristics consisting of opioid prescribing rates (retail opioid prescriptions dispensed per 100 persons per year) and drug poisoning mortality rates. Year fixed effects are added to capture variations such as changes at the federal or nation level, which may have affected states' online SUDT job vacancies equally across states. State fixed effects are included to correct for unobserved heterogeneity. In particular, this two-way fixed effect model (DD approach) allows us to control for all omitted state-specific time-invariant covariates that cause some states to have more job postings related to SUDT than others. Since observations in the same state may have correlated errors, we cluster-correct the standard errors at the state-level. 
Visual inspection of our outcome distribution demonstrates a strong positive skewness, suggesting excess zeros may be an issue. The histogram of online job vacancies in \nameref{S1_Fig} suggests that there might be excess zeros in the vacancy data (63 state-year observations, accounted for 13.7\% of all observations). In addition, only 33\% of states had at least one job ad for therapists. The variance of the count outcome is much larger than the mean (5,531 vs. 108.6). Due to the positive skewness in the distribution of the outcome variables with excess zeros (especially, job postings for each occupation), we also estimated the two-way fixed effects models in two separate estimations for job postings of different occupations. There, the first model is an OLS regression predicting whether a state had any job vacancy in the relevant category. The second model is an OLS estimation which regresses the number of job postings per 100,000 state residents (which takes logged form) on aforementioned predictors. The two-part regression models have applied to both continuous and count data with excess zeros in econometric analyses since 1980s~\cite{Olsen2001,Duan1983}. In our study, the two models were fit separately with standard linear regression software (Stata) instead of using zero-inflated Poisson (or Negative Binomial) to keep the two-way fixed effects setting. 
In order to evaluate the underlying assumption of the DD design in this current study, that in the absence of Medicaid expansion, there would have been parallel trends in the control and treatment states, we present event study results. This helps evaluate whether Medicaid expansion states trends were similar to non-expansion states prior to expansion implementation. In particular, we regress the number of vacancies on dummies for any pre-policy trend periods (4 years or more before expansion, 3 years before expansion, and 2 years before expansion) and dummies for any post periods (implementation year, and 1 or more years after expansion). A significant coefficient of any pre-policy trend periods may suggest a violation of the assumption underlying our DD. We use the same sets of covariates of the DD models in these event study analyses.

\section*{Results}
\begin{figure}[H]
\includegraphics[width=5in]{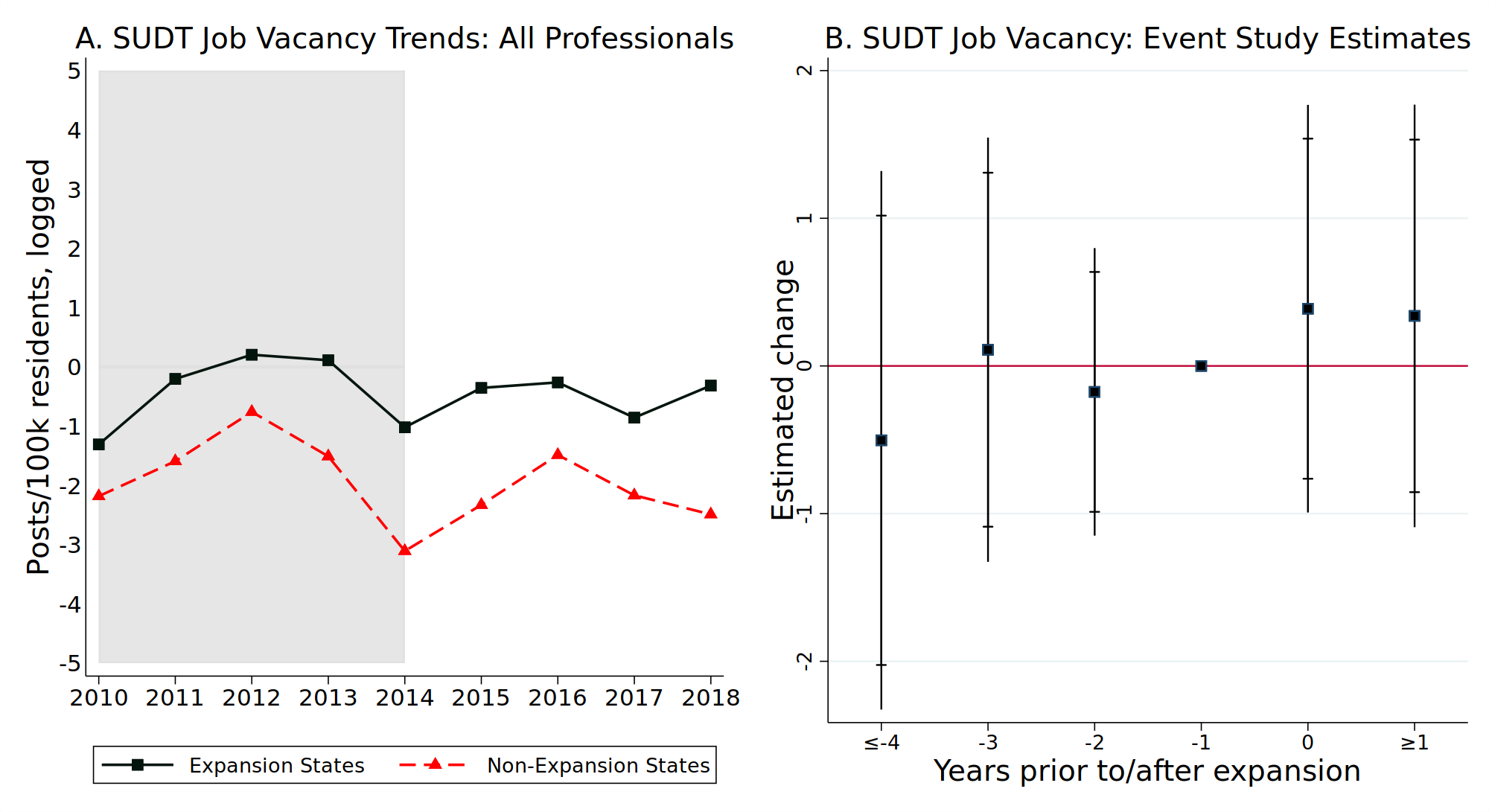}
\caption{{\bf Raw Trends and Event Study Estimates of Job Vacancies for SUDT.}
Authors' calculations based on NAICS-state data from BGT, 2010-18, CDC prescribing rates, CDC drug poisoning mortality rates, and socio-demographic data from the BLS and Census Bureau. Panel A: we calculated the raw means of job postings per 100,000 state residents (which took log forms) for Expansion States and Non-Expansion States from state BGT data. Late expansion states (AK, IN, LA, NH, MI, MT, and PA) are excluded from this comparison. Panel B: plots the estimated difference and its 95 and 90 (bar) percent confidence intervals for each period prior to and after the implementation of Medicaid Expansion. The dependent variable is the logged number of job postings per 100,000 state residents. Late expansion states, together with 43 other states, were included in this analysis. In this event study regression, we controlled for state fixed effects, year fixed effects, median income (logged), unemployment rate, opioid prescribing rates, and age-adjusted mortality rates for drug poisoning (one- year lag values of these control variables).}
\label{fig3}
\end{figure}

Fig~\ref{fig3} shows the raw unadjusted job vacancy trends for the SUDT sector for Medicaid expansion states and non-expansion states (Fig~\ref{fig3}A). We observe fairly consistent patterns between the two sets of states in the pre reform (2014) as well as in the post reform period, which suggests that Medicaid expansion did not cause a meaningful change in SUDT sector hiring attempts. In order to understand the context better, we also examine the pattern of results for all other industries through further explorations. \nameref{S2_Fig}, for all of the industries in the economy (except SUDT sector) and for all healthcare industry (except SUDT sector), exhibits similar parallel patterns pre and post Medicaid expansion. In particular, in \nameref{S2_Fig} we see parallel trends in all non-SUDT industries and healthcare sector prior to as well as after Medicaid expansion.  \nameref{S3_Fig}(A-H) looks at specific occupations next. This analysis shows first the aggregate effect across all occupations, which has a point estimate very close to zero. However, there is evidence of statistically significant increases in some of the separate occupations we study: counselors, entry-level practitioners, and advanced-level primary practitioners. These comparisons provide some preliminary evidence on increases in job vacancies only for certain professions following Medicaid expansion but not others, which leads to an aggregate result of no increase on average~\cite{Mojtabai2019}.

\begin{figure}[H]
\includegraphics[width=5in]{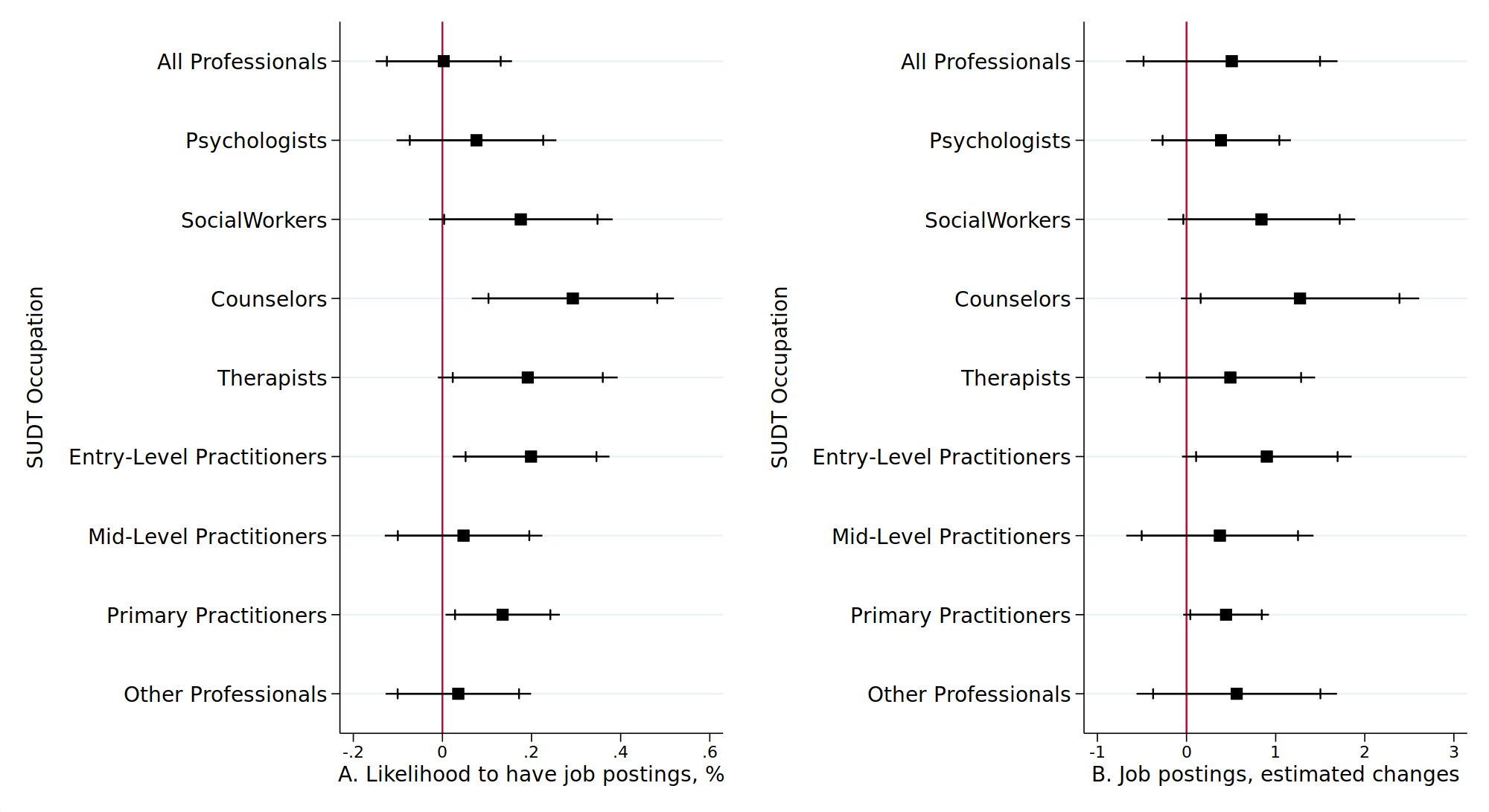}
\caption{{\bf DD Estimates for Impact of Medicaid Expansion on SUDT Job Vacancies by Occupation.}
Authors' estimations based on NAICS-state data from BGT, 2010-18, CDC prescribing rates, CDC mortality rates, and socio-demographic data from the BLS and Census Bureau. Late expansion states (AK, IN, LA, NH, MI, MT, and PA), together with 43 other states, were included in this analysis. We use 1 year lagged values of the control variables. Standard errors were clustered at the state-level. Panel 1: the dependent variable is a binary indicator of whether a state had at least one job vacancy for a specific profession in a certain year. Panel 2: the dependent variable is the number of job vacancies per 100,000 state residents, which takes logged form. A small amount (0.001) was added to this outcome in order to retain zeros in these analyses.}
\label{fig4}
\end{figure}

\begin{figure}[!ht] 
\paragraph*{S1 Fig.}
\includegraphics[width=5in]{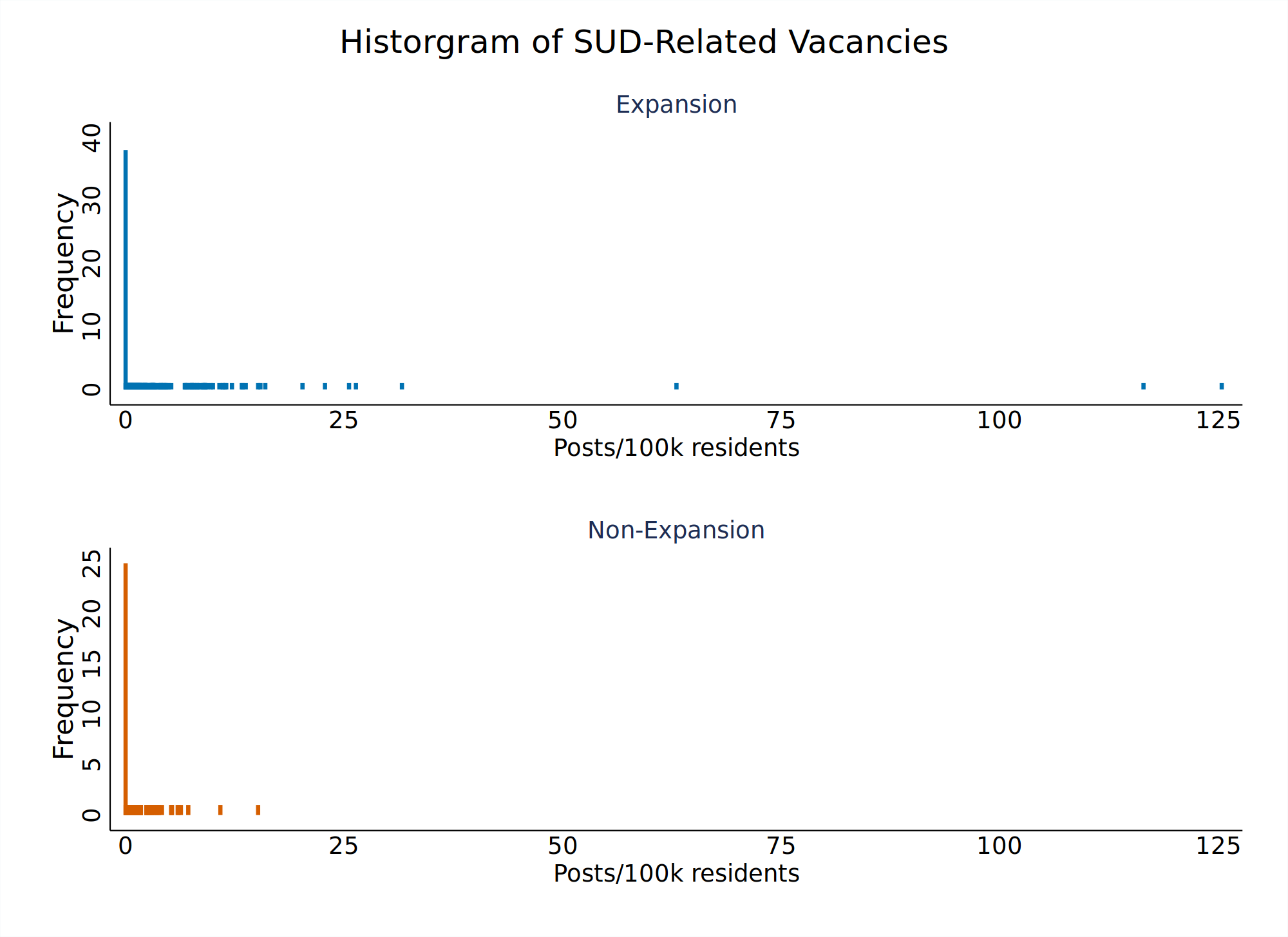}
\label{S1_Fig}
{\bf Histogram of Job Postings in SUDT Sector.} Authors' calculations based on BGT, 2010-18.
\end{figure}

\begin{figure}[!ht] 
\paragraph*{S2 Fig.}
\includegraphics[width=5in]{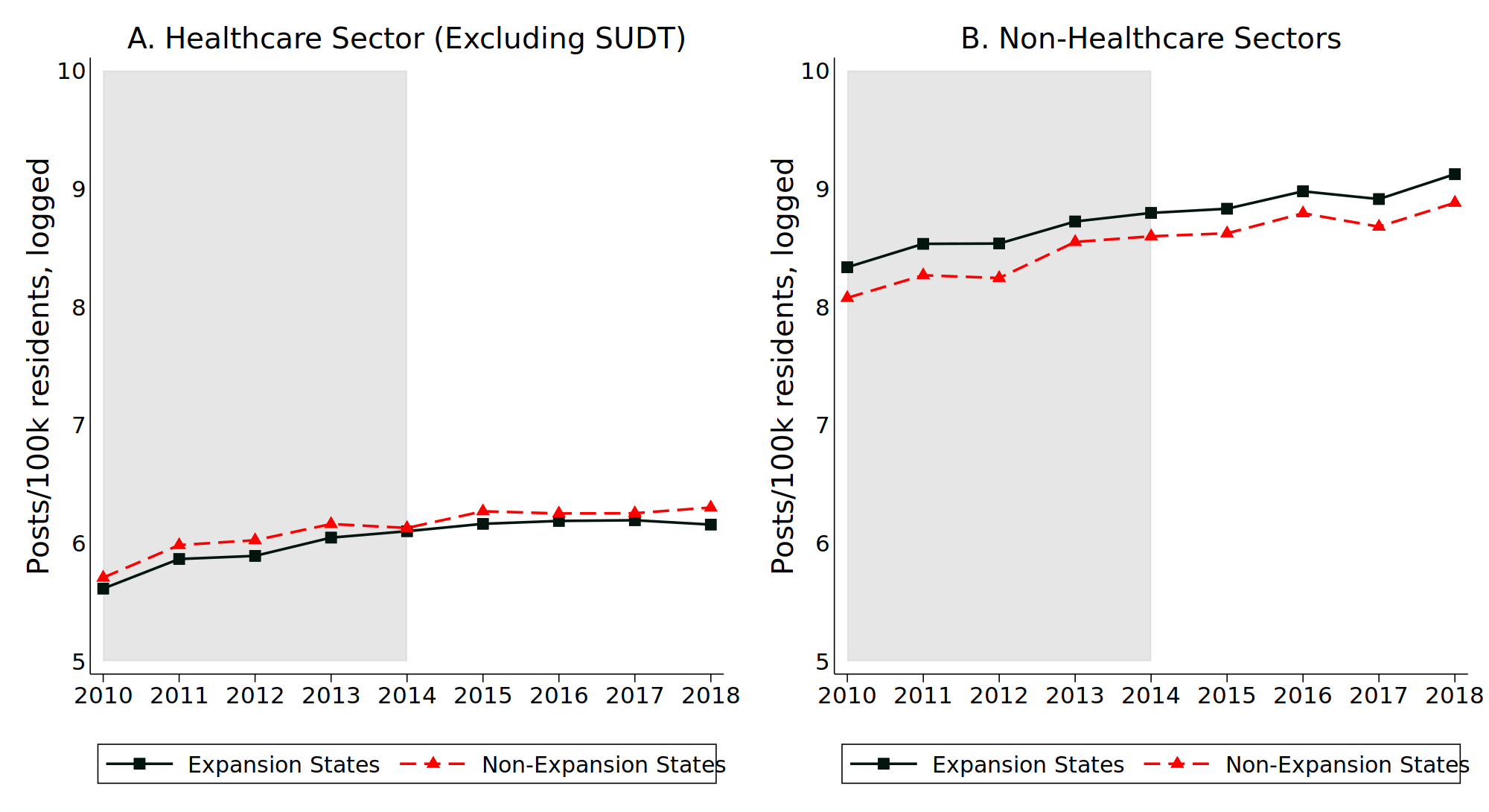}
\label{S2_Fig}
{\bf Job Vacancy Trends for Healthcare and Non-Healthcare Sectors.} Authors' calculations based on NAICS-state data from Burning Glass, 2010-18. In particular, we used the NAICS-state data to show state level and estimated the means for Expansion States and Non-Expansion States. Estimates were adjusted by state populations. Late expansion states (AK, IN, LA, NH, MI, MT, and PA) were excluded from the calculations. 
\end{figure}

\begin{figure}[!ht] 
\paragraph*{S3 Fig.}
\includegraphics[width=5in]{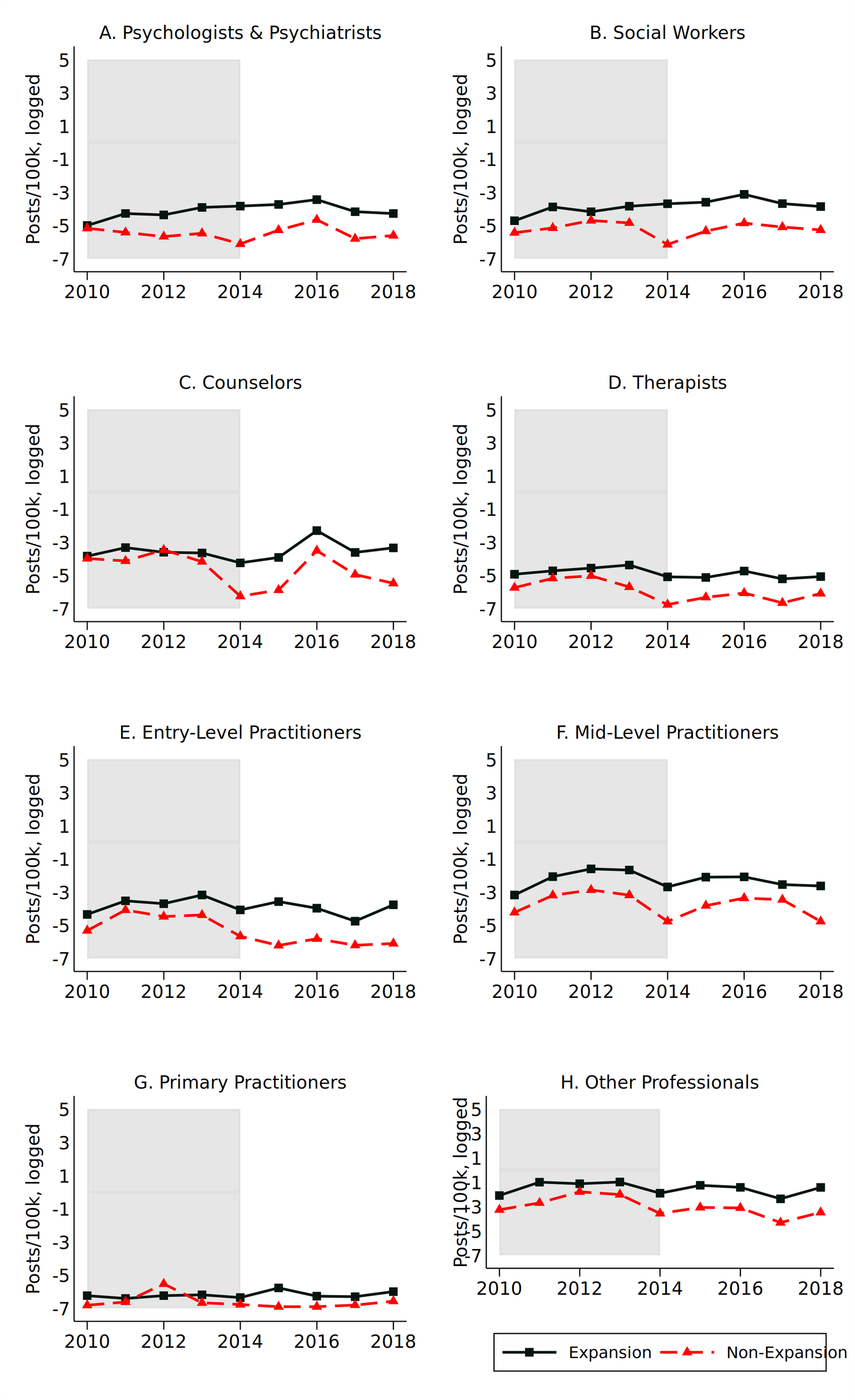}
\label{S3_Fig}
{\bf Job Vacancy Trends for SUDT Occupations for Medicaid Expansion and Non-Expansion States.} Authors' calculations based on NAICS-state data from Burning Glass, 2010-18. In particular, we used the NAICS-state data to show state level and estimated the means for Expansion States and Non-Expansion States. Estimates were adjusted by state populations. Late expansion states (AK, IN, LA, NH, MI, MT, and PA) were excluded from the calculations.
\end{figure}

A key identifying assumption of our DD model was that expansion and non-expansion states would have trended similarly in the absence of expansion. We first visually assessed trends in Fig~\ref{fig3}A. We then formally tested for pre-policy parallel trends by examining the coefficients on the pre-expansion interaction terms in our event study model, presented in Fig~\ref{fig3}B. The coefficients and 95 (and 90) percent confidence intervals for each interaction term are plotted in Fig~\ref{fig4}. This event study analysis suggests that expansion and non-expansion states were similar regarding the frequency of SUDT job ads. Using the DD and event study design, however, we are unable to reject that Medicaid expansion did not increase the number of SUDT job ads. 

We further estimate DD models separately for 7 most relevant SUDT occupations and other occupations: (1) psychiatrists and psychologists; (2) social workers (SOC codes: Mental Health and Substance Abuse Social Workers; Child, Family, and School Social Workers; and Healthcare Social Workers, All Other); (3) counselors (SOC codes: Mental Health Counselors; Educational, Guidance, School, and Vocational Counselors; Substance Abuse and Behavioral Disorder Counselors; and Counselors, All Other); (4) Marriage and Family Therapists; Physical Therapists; Occupational Therapists; Recreational Therapists; Respiratory Therapists; Radiation Therapists; Massage Therapists; (5) Entry-level positions (SOC codes: Home Health Aides; Psychiatric Aides; Physical Therapist Aides; Pharmacy Aides; Personal Care Aides; Medical Assistants, Nursing Assistants, Therapy/Therapist Assistants, Social and Human Service Assistants; Residential Advisors; Technicians; Childcare Workers; Medical Secretaries; and Healthcare Support Workers); (6) Mid-level practitioners (SOC codes: Physician Assistants; Nurse Practitioners, Registered Nurses, Licensed Practical and Licensed Vocational Nurses, Clinical Laboratory Technicians/Technologists, Health Technologists and Technicians); (7) Advance-level primary practitioners (SOD codes: Physicians and Surgeons; Family and General Practitioners; and Health Diagnosing and Treating Practitioners); and (8) all other occupations.

Overall, the DD estimates in Fig~\ref{fig4} suggest that Medicaid expansion does not increase the likelihood of attempt to hire SUDT professionals or the number of job postings (all professionals, collectively), although (perhaps unsuprisingly given the number of different specifications we run), there are some coefficients that are statistically significant for specific occupations. The categories with statistically significant coefficients are counselors (increase by 29\% (p$<$0.05)), entry-level practitioners (increase by 20\% (p$<$0.05)), and primary practitioners (increase by 13.5\% (p$<$0.05)), compared to non-expansion states. The unreported event study results suggest that expansion and non-expansion states occupational demand would have trended similarly in the absence of expansion, except for the likelihood of hiring social workers. The DD estimates in Fig~\ref{fig4}B show no coefficients that are statistically significant at conventional levels, suggesting no detectable effects of Medicaid on the number of hiring attempts on aggregate or for specific occupations. \nameref{S1_Table} provides details of these DD estimations. There are some statistically significant effects: after accounting for all state-specific time-invariant characteristics and changes at the national level through state fixed effects and year fixed effects, drug poisoning mortality rate is positively associated with hiring attempts for psychologists and psychiatrists (Model 1, ~\nameref{S1_Table}). We also find that the opioid prescribing rate is positively correlated with the likelihood to attempt to hire entry-level practitioners (Model 5, ~\nameref{S1_Table}). On the whole, these regression estimates however do not indicate substantial impacts due to Medicaid expansion.

\section*{Discussion}

This paper provides the first analyses of the workforce demand side in the SUDT sector and how public insurance expansion is associated with growth. First, we note that the SUDT sector measured here as mental health and SUDT is about 5\% of all healthcare sector hiring attempts 
in the US and that this has not increased substantially over time. The lack of overall growth of SUDT job demand is unexpected, given that SAMHSA predicted increased demand for SUDT medical use and services~\cite{SubstanceAbuseandMentalHealthServicesAdministration2018a,Hoge2013}. The dramatic increases in demand for SUDT services and inadequate behavioral health workforce had been predicted following major health care reforms such as the ACA by policy makers and independent experts.~\cite{Hoge2013,Dall2013} Even though existing evidence shows a recent increase in SUD medication access, it is yet unknown whether Medicaid expansion has also led to a growth in hiring attempts in the SUDT workforce, partly due to data scarcity~\cite{SubstanceAbuseandMentalHealthServicesAdministration2018a}. 
Use a novel data source covering virtually all U.S. online job postings by employers hiring in the SUDT workforce, we studied hiring trends in total and by top occupations within these relevant industries. Comparing the raw trends in SUDT job postings, we do not find that Medicaid expansion is associated with a visually detectable increase in SUDT job postings in the post reform period; there are also fairly consistent patterns between the two sets of states in the pre reform (2014). Applying a DD design in the two-way fixed effect models, we are unable to reject that Medicaid expansion did not increase the number of SUDT job ads as a whole, although some statistically significant coefficient emerge when looking at each occupation level separately.

This study has several limitations. First, as mentioned throughout, standard labor data classifications, which use the NAICS, only allow examination of the mental health and SUDT sector combined. However, we argue that how Medicaid expansion affects behavioral health workforce is extremely relevant for SUD given strong comorbidity patterns. However, since there is great need to understand the resource available for substance addictions treatment specifically, this should be seen as a major limitation of our research. Second, the findings only speak to hiring attempts: when data on actual filling of posts are released for more recent years, research should examine the effects on the actual stock of employees, as our ultimate interest is in assessing adequacy of the SUDT workforce. Nevertheless, these findings are particularly relevant as some states consider changing their public insurance programs through implementation of Medicaid work requirements and other changes to the accessibility of the program, and to states that have yet to expand Medicaid. Third, several studies pointed out that online advertisements often target high-skill technical and managerial candidates, whereas blue-collar occupations are advertised off-line.~\cite{Carnevale2014} Additionally, online job postings may over-represent growing firms~\cite{Hershbein2018}. 

The SUDT related labor force is comprised of occupations with a variety of skill levels within clinical settings such as primary care, behavioral care, or integrated care~\cite{SubstanceAbuseandMentalHealthServicesAdministration2018a}. Entry-level positions such as personal care assistants and nurse assistants often require relatively little prior training, positions for physicians and psychologists/psychiatrists require most advanced degrees (doctorate), and physician assistants, nurse practitioners, registered nurses, and clinical laboratory technicians/technologists require master’s or bachelor’s degree~\cite{Frogner2013}. Therefore, this study further examined the effects of Medicaid expansion on reshaping the composition of the SUDT related workforce. During 2010-2018, most SUDT related hiring attempts had been made for registered nurses, medical and health service managers, mental health counselors, personal care aides, and nurse practitioners. The DD estimates suggest that expansion states are more likely to hire counselors, social workers, entry-level practitioners, as well as primary care practitioners. These findings are suggestive of compositional changes that may have clinical repercussions. These represent fruitful areas for future research to complement findings of increased use of treatment medication~\cite{Wen2017,Saloner2018,Maclean2019}. The increased hiring attempts for both behavioral health and primary care practitioners may suggest that SUDT related establishments are recruiting a diverse workforce and integrating primary and behavioral health care. Despite prior projections that every 10 percent increase in the demand for SUD related treatment would result in the need for 6,800 additional SUD related counselors~\cite{Hoge2013}, our results suggest that although the mortality consequences of the opioid crisis continued to mount during our study period, the treatment workforce hiring attempts failed to show substantial increases; future research should continue to examine impact of alternate policy levers to provide a more comprehensive body of knowledge regarding factors that could expand the availability of treatment.

\section*{Acknowledgments}
The authors would like to thank Burning Glass Technologies (BGT) for data access. The authors also thank Livia Crim and Anurag Joshi for excellent research assistance. We are grateful to Jason Turi and Amanda Abraham for comments.  

\nolinenumbers

%

\section*{Supporting information}


\begin{table}[!ht]
\paragraph*{S1 Table.}
\label{S1_Table}
{\bf DD Estimates for Impact of Medicaid Expansion on Job Vacancies of SUD-Related Industries} Authors' estimations based on NAICS-state data from BGT, 2010-18, CDC prescribing rates, CDC drug poisoning mortality rates, and socio-demographic data from the BLS and Census Bureau. In these DD regressions, we used OLS estimations and controlled for state fixed effects, year fixed effects, median income (logged), unemployment rate, opioid prescribing rates, and age-adjusted mortality rates for drug poisoning. We used 1 year lagged values of the control variables. Panel 1: the dependent variable is a binary indicator of whether or not a state had at least one job vacancy for a specific professional in a certain year. Panel 2: the dependent variable is the number of job vacancies per 100,000 state residents, which takes logged form. A small amount (0.001) was added to this outcome in order to remain zeros in these analyses. * p$<$0.1 ** p$<$0.05 *** p$<$0.01.

\begin{adjustwidth}{-2.5in}{.25in} 
\centering 
\begin{tabular}{lccccccc}
 \hline 

 
              &\multicolumn{1}{c}{Model 1}   &\multicolumn{1}{c}{Model 2}   &\multicolumn{1}{c}{Model 3}   &\multicolumn{1}{c}{Model 4} &\multicolumn{1}{c}{Model 5}   &\multicolumn{1}{c}{Model 6}   &\multicolumn{1}{c}{Model 7}   \\   
               &\multicolumn{1}{c}{Psychologists/}   &\multicolumn{1}{c}{Social}   &\multicolumn{1}{c}{Counselors}   &\multicolumn{1}{c}{Therapists} &\multicolumn{1}{c}{Entry-Level}   &\multicolumn{1}{c}{Mid-Level}   &\multicolumn{1}{c}{Primary}   \\   
		&\multicolumn{1}{c}{Psychiatrists}   &\multicolumn{1}{c}{workers}   &\multicolumn{1}{c}{}   &\multicolumn{1}{c}{} &\multicolumn{1}{c}{Practitioners}   &\multicolumn{1}{c}{Practitioners}   &\multicolumn{1}{c}{Practitioners}   \\   

\hline
\multicolumn{8}{l}{Panel 1: Y - Having Any Job Postings} \\
Expansion$\times$Post-2014&    0.076   &     0.18*  &     0.29** &     0.19*  &     0.20** &    0.047   &     0.14** \\
                &  (0.089)   &   (0.10)   &   (0.11)   &   (0.10)   &  (0.088)   &  (0.088)   &  (0.064)   \\
Unemployment rates, \%&   -0.037   &   -0.061*  &   -0.014   &   -0.016   &    0.013   &  -0.0072   &   -0.038** \\
                &  (0.024)   &  (0.033)   &  (0.031)   &  (0.025)   &  (0.026)   &  (0.028)   &  (0.018)   \\
Median income, logged&     0.63   &    -1.93   &    -1.04   &     0.61   &    0.094   &    -1.16   &     0.61   \\
                &   (1.00)   &   (1.19)   &   (1.22)   &   (0.89)   &   (0.79)   &   (1.17)   &   (0.68)   \\
Opioid prescribing rates&   0.0057   &  -0.0044   &   0.0039   &  0.00072   &   0.0090** &   0.0021   &  -0.0041   \\
                & (0.0064)   & (0.0048)   & (0.0064)   & (0.0042)   & (0.0035)   & (0.0060)   & (0.0042)   \\
Drug poisoning death rates&   0.0089*  &  -0.0024   &  -0.0016   &  -0.0082   &   0.0031   &   0.0028   &   0.0021   \\
                & (0.0047)   & (0.0047)   & (0.0070)   & (0.0055)   & (0.0054)   & (0.0063)   & (0.0041)   \\

    \hline
	Dep. Variable Mean&\multicolumn{1}{c}{0.45}   &\multicolumn{1}{c}{0.51}   &\multicolumn{1}{c}{0.54}   &\multicolumn{1}{c}{0.33}   &\multicolumn{1}{c}{0.42}   &\multicolumn{1}{c}{0.71}   &\multicolumn{1}{c}{0.13}   \\
  Adj R-squared   &      \multicolumn{1}{c}{0.491}      &   \multicolumn{1}{c}{0.464}         &      \multicolumn{1}{c}{0.509}      &     \multicolumn{1}{c}{0.521}       &    \multicolumn{1}{c}{0.519}        &      \multicolumn{1}{c}{0.362}      &      \multicolumn{1}{c}{0.317}      \\
               \hline
             \multicolumn{8}{l}{Panel 2: Y - Number of Postings per 100,000 State Residents, Logged} \\
Expansion$\times$Post-2014&     0.39   &     0.84   &     1.27*  &     0.49   &     0.90*  &     0.37   &     0.44*  \\
                &   (0.39)   &   (0.52)   &   (0.67)   &   (0.47)   &   (0.47)   &   (0.52)   &   (0.24)   \\
Unemployment rates, \%&    -0.11   &    -0.26*  &    -0.12   &   -0.094   &  -0.0093   &   -0.093   &    -0.12*  \\
                &   (0.12)   &   (0.13)   &   (0.16)   &   (0.11)   &   (0.15)   &   (0.15)   &  (0.062)   \\
Median income, logged&     2.66   &    -6.31   &    -2.42   &     4.08   &     2.18   &    -4.88   &     2.94   \\
                &   (5.84)   &   (5.72)   &   (6.64)   &   (4.51)   &   (4.61)   &   (6.62)   &   (2.69)   \\
Opioid prescribing rates&    0.030   &   -0.029   &    0.015   &  -0.0044   &    0.032   &    0.022   &   -0.022   \\
                &  (0.034)   &  (0.023)   &  (0.035)   &  (0.021)   &  (0.022)   &  (0.035)   &  (0.022)   \\
Drug poisoning death rates&    0.064** &   -0.029   &   -0.022   &   -0.045   &  -0.0039   &    0.022   &  -0.0028   \\
                &  (0.026)   &  (0.026)   &  (0.041)   &  (0.028)   &  (0.030)   &  (0.037)   &  (0.015)   \\

    \hline
     Dep. Variable Mean&\multicolumn{1}{c}{0.137}   &\multicolumn{1}{c}{0.196}   &\multicolumn{1}{c}{0.222}   &\multicolumn{1}{c}{0.061}   &\multicolumn{1}{c}{0.423}   &\multicolumn{1}{c}{0.479}   &\multicolumn{1}{c}{0.013}   \\
  Adj R-squared   &      \multicolumn{1}{c}{0.614}      &   \multicolumn{1}{c}{0.531}         &      \multicolumn{1}{c}{0.536}      &     \multicolumn{1}{c}{0.556}       &    \multicolumn{1}{c}{0.621}        &      \multicolumn{1}{c}{0.539}      &      \multicolumn{1}{c}{0.392}      \\
\hline
\end{tabular}
\end{adjustwidth}
\end{table}

\end{document}